\documentclass[bibyear]{aa}
\usepackage{graphicx}
\usepackage[varg]{txfonts}
\usepackage{color}
\begin{document}

\title{Alignment of quasar polarizations with large-scale
structures\thanks{Based on observations made with ESO Telescopes at
the La Silla Paranal Observatory under programme ID 092.A-0221}}

\author{D. Hutsem\'ekers\inst{1}, 
        L. Braibant\inst{1},
        V. Pelgrims\inst{1},
        D. Sluse\inst{2}
        }

\institute{
    Institut d'Astrophysique et de G\'eophysique,
    Universit\'e de Li\`ege, All\'ee du 6 Ao\^ut 17, B5c, B-4000
    Li\`ege, Belgium
    \and 
    Argelander-Institut für Astronomie, Auf dem Hügel 71, 53121 Bonn, Germany
    }
  
\date{Received ; accepted: }

\titlerunning{Alignment of quasar polarizations with large-scale
structures} 
   
\authorrunning{Hutsem\'ekers D. et al.}

\abstract{We have measured the optical linear polarization of quasars
belonging to Gpc-scale quasar groups at redshift $z \sim$ 1.3.  Out of
93 quasars observed, 19 are significantly polarized.  We found that
quasar polarization vectors are either parallel or perpendicular to
the directions of the large-scale structures to which they belong.
Statistical tests indicate that the probability that this effect can
be attributed to randomly oriented polarization vectors is of the
order of 1\%.  We also found that quasars with polarization
perpendicular to the host structure preferentially have large emission
line widths while objects with polarization parallel to the host
structure preferentially have small emission line widths.  Considering
that quasar polarization is usually either parallel or perpendicular
to the accretion disk axis depending on the inclination with respect
to the line of sight, and that broader emission lines originate from
quasars seen at higher inclinations, we conclude that quasar spin axes
are likely parallel to their host large-scale structures.}


\maketitle
%
%________________________________________________________________

\section{Introduction}
\label{sec:intro}

Hutsem\'ekers et al. (\cite{hut98, hut01, hut05}, hereafter Papers I,
II, III) have reported alignments of quasar optical linear
polarizations extending over Gpc-scale regions of the sky at redshift
$z \sim 1$ (see also Jain et al. \cite{jai04}, Shurtleff \cite{shu13},
Pelgrims and Cudell \cite{pel14}).  Possible effects modifying the
polarization of light along the line of sight, in particular mixing
with axion-like particles, have been investigated in details (e.g.,
Das et al. \cite{das05}, Agarwal et al. \cite{aga12}).  However, due
to the absence of comparable circular polarization, these mechanisms
have been essentially ruled out (Hutsem\'ekers et al. \cite{hut10},
Payez et al. \cite{pay11}).

Since quasar polarization is often related to the object geometry,
another interpretation would be that quasars themselves are aligned,
presumably with the structure to which they belong. To test this
hypothesis, we have measured the polarization of quasars belonging to
the large quasar group (LQG) constituted of the groups U1.27 (aka
Huge-LQG) and U1.28 (aka CCLQG) described in Clowes et
al. (\cite{clo13}). These quasar structures extend over Gpc scales at
redshift $z \sim 1.3$, possibly beyond the homogeneity scale of the
concordance cosmology ($\sim$ 370 Mpc, Yadav et al. \cite{yad10}; but
see Nadathur \cite{nad13}, Einasto et al. \cite{ein14}).

\section{Observations and polarization  measurements}
\label{sec:obs}

Observations were carried out at the European Southern Observatory,
Paranal, on March 22-26, 2014, using the Very Large Telescope equipped
with the FORS2 intrument in the standard imaging polarimetry mode
IPOL\footnote{FORS User Manual, VLT-MAN-ESO-13100-1543, Issue 92.0}.
Linear polarimetry is performed by inserting in the parallel beam a
Wollaston prism which splits the incoming light rays into two
orthogonally polarized beams separated by 22\arcsec .  Image
overlapping is avoided by inserting a special mask in the focal
plane. To measure the normalized Stokes parameters $q$ and $u$, 4
frames are obtained with the half-wave plate rotated at 4 position
angles, 0\degr, 22.5\degr, 45\degr, and 67.5\degr . This procedure
allows us to remove most of the instrumental polarization. The linear
polarization degree $p$ and position angle $\theta$ are derived using
$p = (q^2 + u^2)^{1/2}$ and $\theta = 1/2 \arctan \, (u/q)$ so that $q
= p \cos 2\theta$ and $u = p \sin 2\theta$.  Since orthogonally
polarized images of the object are simultaneously recorded, the
measured polarization does not depend on variable transparency or
seeing.

\onltab{
\begin{longtab}
\onecolumn 
\begin{longtable}{lcccrrrrrrrr}
\caption{\label{poltab} The linear polarization of 93 quasars}\\
\hline\hline
Object & $z$ & LQG & Date & $q$ & $u$ & $p$ & $\sigma_{p}$ & $p_0$ & $\theta$ \ & $\sigma_{\theta}$ & Notes\\ 
& & & (dd/mm/yy) & (\%) & (\%) & (\%) & (\%) & (\%) &  ($\degr$) & ($\degr$) & \\
\hline
\endfirsthead
\caption{continued.}\\
\hline\hline
Object & $z$ & LQG & Date & $q$ & $u$ & $p$ & $\sigma_{p}$ & $p_0$ & $\theta$ \ & $\sigma_{\theta}$ & Notes\\ 
& & & (dd/mm/yy) & (\%) & (\%) & (\%) & (\%) & (\%) &  ($\degr$) & ($\degr$) & \\
\hline
\endhead
\hline
\endfoot
SDSSJ104938.22+214829.3 & 1.2352 & 1 & 25/03/14 &  -0.02 &   0.00 &   0.02 &  0.07 &  0.00 &    $-$ &    $-$ & \\
SDSSJ105140.40+203921.1 & 1.1742 & 1 & 22/03/14 &  -0.01 &  -0.14 &   0.14 &  0.06 &  0.13 &    133 &     13 & \\
SDSSJ105224.08+204634.1 & 1.2032 & 1 & 23/03/14 &  -0.18 &  -0.02 &   0.18 &  0.10 &  0.16 &     93 &     18 & \\
SDSSJ105258.16+201705.4 & 1.2526 & 1 & 23/03/14 &  -0.09 &   0.24 &   0.26 &  0.21 &  0.19 &     55 &     32 & \\
SDSSJ105421.90+212131.2 & 1.2573 & 1 & 23/03/14 &  -1.04 &  -0.09 &   1.04 &  0.08 &  1.04 &     93 &      2 & \\
SDSSJ105446.73+195710.5 & 1.2195 & 1 & 22/03/14 &  -1.64 &   0.93 &   1.89 &  0.23 &  1.88 &     75 &      4 & \\
SDSSJ105525.18+191756.3 & 1.2005 & 1 & 23/03/14 &  -0.16 &   0.34 &   0.38 &  0.13 &  0.36 &     58 &     10 & \\
SDSSJ105556.22+184718.4 & 1.2767 & 1 & 23/03/14 &   0.08 &   0.02 &   0.08 &  0.12 &  0.00 &    $-$ &    $-$ & \\
SDSSJ105611.27+170827.5 & 1.3316 & 1 & 25/03/14 &   0.01 &   1.29 &   1.29 &  0.08 &  1.29 &     45 &      2 & \\
SDSSJ105714.02+184753.3 & 1.2852 & 1 & 24/03/14 &  -0.02 &  -0.05 &   0.05 &  0.09 &  0.00 &    $-$ &    $-$ & \\
SDSSJ105805.09+200341.0 & 1.2731 & 1 & 23/03/14 &   0.13 &   0.17 &   0.21 &  0.09 &  0.19 &     26 &     13 & \\
SDSSJ105832.01+170456.0 & 1.2813 & 1 & 22/03/14 &  -0.35 &  -0.09 &   0.36 &  0.16 &  0.33 &     97 &     14 & \\
SDSSJ105840.49+175415.5 & 1.2687 & 1 & 24/03/14 &   0.14 &  -0.24 &   0.28 &  0.11 &  0.26 &    150 &     12 & \\
SDSSJ105928.57+164657.9 & 1.2993 & 1 & 24/03/14 &   0.02 &   0.24 &   0.24 &  0.15 &  0.20 &     43 &     21 & \\
SDSSJ110016.88+193624.7 & 1.2399 & 1 & 22/03/14 &   0.88 &  -0.72 &   1.14 &  0.23 &  1.12 &    160 &      6 & \\
SDSSJ110039.99+165710.3 & 1.2997 & 1 & 22/03/14 &  -0.06 &   0.16 &   0.17 &  0.12 &  0.14 &     56 &     25 & \\
SDSSJ104139.15+143530.2 & 1.2164 & 2 & 22/03/14 &  -0.21 &   0.14 &   0.25 &  0.16 &  0.21 &     73 &     22 & \\
SDSSJ104321.62+143600.2 & 1.2660 & 2 & 24/03/14 &  -0.19 &  -0.11 &   0.22 &  0.10 &  0.20 &    105 &     14 & \\
SDSSJ104430.92+160245.0 & 1.2294 & 2 & 22/03/14 &  -0.08 &   0.04 &   0.09 &  0.07 &  0.07 &     76 &     30 & \\
SDSSJ104445.03+151901.6 & 1.2336 & 2 & 23/03/14 &   1.13 &  -0.53 &   1.25 &  0.11 &  1.25 &    168 &      3 & \\
SDSSJ104520.62+141724.2 & 1.2650 & 2 & 23/03/14 &   0.19 &   0.01 &   0.19 &  0.12 &  0.16 &      1 &     22 & \\
SDSSJ104604.05+140241.2 & 1.2884 & 2 & 22/03/14 &  -0.17 &   0.50 &   0.53 &  0.15 &  0.51 &     54 &      8 & \\
SDSSJ104616.31+164512.6 & 1.2815 & 2 & 24/03/14 &  -1.24 &   0.13 &   1.25 &  0.11 &  1.25 &     87 &      3 & \\
SDSSJ104624.25+143009.1 & 1.3620 & 2 & 23/03/14 &  -0.04 &  -0.20 &   0.20 &  0.10 &  0.18 &    129 &     16 & \\
SDSSJ104813.63+162849.1 & 1.2905 & 2 & 22/03/14 &  -0.04 &   0.16 &   0.17 &  0.12 &  0.14 &     53 &     25 & \\
SDSSJ104859.74+125322.3 & 1.3597 & 2 & 25/03/14 &  -0.02 &   0.72 &   0.72 &  0.13 &  0.71 &     46 &      5 & \\
SDSSJ104915.66+165217.4 & 1.3459 & 2 & 23/03/14 &  -0.14 &  -0.51 &   0.53 &  0.09 &  0.52 &    127 &      5 & \\
SDSSJ104922.60+154336.1 & 1.2590 & 2 & 22/03/14 &  -0.26 &   0.03 &   0.26 &  0.15 &  0.22 &     87 &     19 & \\
SDSSJ104924.30+154156.0 & 1.2965 & 2 & 25/03/14 &  -0.14 &   0.16 &   0.21 &  0.12 &  0.18 &     66 &     19 & \\
SDSSJ104941.67+151824.6 & 1.3390 & 2 & 24/03/14 &   0.51 &  -1.21 &   1.31 &  0.13 &  1.30 &    146 &      3 & \\
SDSSJ104947.77+162216.6 & 1.2966 & 2 & 22/03/14 &   0.13 &  -0.01 &   0.13 &  0.14 &  0.00 &    $-$ &    $-$ & \\
SDSSJ104954.70+160042.3 & 1.3373 & 2 & 23/03/14 &  -0.21 &  -0.29 &   0.36 &  0.17 &  0.32 &    117 &     15 & \\
SDSSJ105001.22+153354.0 & 1.2500 & 2 & 25/03/14 &  -0.03 &  -0.02 &   0.03 &  0.11 &  0.00 &    $-$ &    $-$ & \\
SDSSJ105042.26+160056.0 & 1.2591 & 2 & 25/03/14 &   0.01 &   0.20 &   0.20 &  0.08 &  0.18 &     44 &     12 & \\
SDSSJ105104.16+161900.9 & 1.2502 & 2 & 22/03/14 &   0.24 &  -0.01 &   0.24 &  0.10 &  0.22 &    179 &     13 & \\
SDSSJ105117.00+131136.0 & 1.3346 & 2 & 23/03/14 &  -0.37 &   0.46 &   0.59 &  0.22 &  0.55 &     64 &     11 & \\
SDSSJ105119.60+142611.4 & 1.3093 & 2 & 25/03/14 &   0.18 &   0.03 &   0.18 &  0.10 &  0.16 &      5 &     18 & \\
SDSSJ105122.98+115852.3 & 1.3085 & 2 & 23/03/14 &   0.01 &  -0.02 &   0.02 &  0.10 &  0.00 &    $-$ &    $-$ & \\
SDSSJ105125.72+124746.3 & 1.2810 & 2 & 25/03/14 &   0.08 &  -0.06 &   0.10 &  0.08 &  0.07 &    160 &     31 & RadioS \\
SDSSJ105132.22+145615.1 & 1.3607 & 2 & 22/03/14 &  -0.40 &   0.15 &   0.43 &  0.10 &  0.42 &     80 &      7 & \\
SDSSJ105144.88+125828.9 & 1.3153 & 2 & 25/03/14 &   0.21 &   0.13 &   0.25 &  0.07 &  0.24 &     15 &      8 & RadioS \\
SDSSJ105210.02+165543.7 & 1.3369 & 2 & 24/03/14 &  -0.16 &  -0.23 &   0.28 &  0.09 &  0.27 &    117 &     10 & \\
SDSSJ105222.13+123054.1 & 1.3162 & 2 & 24/03/14 &  -0.08 &   0.32 &   0.33 &  0.13 &  0.31 &     52 &     12 & \\
SDSSJ105223.68+140525.6 & 1.2483 & 2 & 22/03/14 &   0.27 &   0.41 &   0.49 &  0.11 &  0.48 &     28 &      7 & \\
SDSSJ105245.80+134057.4 & 1.3544 & 2 & 25/03/14 &   0.65 &   1.15 &   1.32 &  0.11 &  1.32 &     30 &      2 & RadioS \\
SDSSJ105257.17+105933.5 & 1.2649 & 2 & 24/03/14 &   0.54 &  -0.15 &   0.56 &  0.11 &  0.55 &    172 &      6 & \\
SDSSJ105412.67+145735.2 & 1.2277 & 2 & 24/03/14 &  -0.18 &  -0.32 &   0.37 &  0.10 &  0.36 &    121 &      8 & RadioS \\
SDSSJ105435.64+101816.3 & 1.2600 & 2 & 24/03/14 &  -0.03 &   0.09 &   0.10 &  0.08 &  0.07 &     55 &     31 & \\
SDSSJ105442.71+104320.6 & 1.3348 & 2 & 24/03/14 &   0.71 &  -0.18 &   0.73 &  0.11 &  0.72 &    173 &      4 & \\
SDSSJ105523.03+130610.7 & 1.3570 & 2 & 25/03/14 &   0.03 &   0.00 &   0.03 &  0.10 &  0.00 &    $-$ &    $-$ & \\
SDSSJ105525.68+113703.0 & 1.2893 & 2 & 24/03/14 &  -0.36 &   2.52 &   2.55 &  0.10 &  2.55 &     49 &      1 & RadioS \\
SDSSJ105541.83+111754.2 & 1.3298 & 2 & 25/03/14 &  -0.45 &   0.01 &   0.45 &  0.11 &  0.44 &     89 &      7 & \\
SDSSJ105621.90+143401.0 & 1.2333 & 2 & 23/03/14 &  -0.42 &   0.13 &   0.44 &  0.13 &  0.42 &     82 &      9 & \\
SDSSJ105637.49+150047.5 & 1.3713 & 2 & 25/03/14 &  -0.37 &   0.46 &   0.59 &  0.16 &  0.57 &     65 &      8 & \\
SDSSJ105637.98+100307.2 & 1.2730 & 2 & 23/03/14 &  -0.08 &   0.16 &   0.18 &  0.12 &  0.15 &     58 &     23 & \\
SDSSJ105655.36+144946.2 & 1.2283 & 2 & 24/03/14 &  -0.14 &  -0.05 &   0.15 &  0.13 &  0.10 &     99 &     39 & \\
SDSSJ105855.33+081350.7 & 1.2450 & 3 & 23/03/14 &  -0.01 &   0.29 &   0.29 &  0.09 &  0.28 &     46 &      9 & \\
SDSSJ110006.02+092638.7 & 1.2485 & 3 & 22/03/14 &   0.17 &   0.14 &   0.22 &  0.09 &  0.20 &     19 &     13 & \\
SDSSJ110148.66+082207.1 & 1.1940 & 3 & 24/03/14 &   0.10 &  -0.19 &   0.22 &  0.14 &  0.18 &    149 &     22 & \\
SDSSJ110217.19+083921.1 & 1.2355 & 3 & 23/03/14 &   0.35 &  -0.02 &   0.35 &  0.12 &  0.33 &    178 &     10 & \\
SDSSJ110504.46+084535.3 & 1.2371 & 3 & 25/03/14 &   0.10 &  -0.10 &   0.14 &  0.10 &  0.11 &    157 &     26 & \\
SDSSJ110621.40+084111.2 & 1.2346 & 3 & 22/03/14 &   0.09 &   0.10 &   0.14 &  0.15 &  0.00 &    $-$ &    $-$ & \\
SDSSJ110736.60+090114.7 & 1.2266 & 3 & 22/03/14 &  -0.42 &   0.08 &   0.43 &  0.11 &  0.42 &     85 &      8 & RadioS \\
SDSSJ110744.61+095526.9 & 1.2228 & 3 & 25/03/14 &   0.44 &   0.25 &   0.51 &  0.08 &  0.50 &     15 &      5 & \\
SDSSJ111007.89+104810.3 & 1.2097 & 3 & 23/03/14 &   0.01 &   0.55 &   0.55 &  0.12 &  0.54 &     45 &      6 & \\
SDSSJ111009.58+075206.8 & 1.2123 & 3 & 24/03/14 &   0.67 &   1.68 &   1.81 &  0.17 &  1.80 &     34 &      3 & \\
SDSSJ111416.17+102327.5 & 1.2053 & 3 & 26/03/14 &   0.26 &   0.02 &   0.26 &  0.08 &  0.25 &      3 &      9 & \\
SDSSJ111545.30+081459.8 & 1.1927 & 3 & 24/03/14 &  -0.30 &   0.16 &   0.34 &  0.14 &  0.31 &     76 &     13 & \\
SDSSJ111802.11+103302.4 & 1.2151 & 3 & 23/03/14 &   1.01 &  -3.84 &   3.97 &  0.10 &  3.97 &    142 &      1 & \\
SDSSJ111823.21+090504.9 & 1.1923 & 3 & 24/03/14 &  -0.10 &  -0.01 &   0.10 &  0.18 &  0.00 &    $-$ &    $-$ & \\
SDSSJ112019.62+085905.1 & 1.2239 & 3 & 23/03/14 &   0.38 &  -0.15 &   0.41 &  0.13 &  0.39 &    170 &     10 & \\
SDSSJ112059.27+101109.2 & 1.2103 & 3 & 24/03/14 &   0.44 &   0.29 &   0.53 &  0.22 &  0.49 &     17 &     13 & \\
SDSSJ112109.76+075958.6 & 1.2369 & 3 & 25/03/14 &  -0.01 &   0.33 &   0.33 &  0.08 &  0.32 &     46 &      7 & \\
SDSSJ104114.06+034312.0 & 1.2633 & 4 & 26/03/14 &  -0.26 &   0.41 &   0.49 &  0.16 &  0.46 &     61 &     10 & BAL \\
SDSSJ104115.58+051345.0 & 1.2553 & 4 & 26/03/14 &  -0.14 &   0.27 &   0.30 &  0.13 &  0.27 &     59 &     14 & \\
SDSSJ104116.79+035511.4 & 1.2444 & 4 & 26/03/14 &  -1.46 &  -0.51 &   1.55 &  0.11 &  1.55 &    100 &      2 & LoBAL \\
SDSSJ104225.63+035539.1 & 1.2293 & 4 & 26/03/14 &   0.48 &   0.50 &   0.69 &  0.08 &  0.69 &     23 &      3 & \\
SDSSJ104256.38+054937.4 & 1.3555 & 4 & 26/03/14 &   0.20 &  -0.36 &   0.41 &  0.12 &  0.39 &    149 &      9 & \\
SDSSJ104425.80+060925.6 & 1.2523 & 4 & 26/03/14 &  -0.36 &  -0.20 &   0.41 &  0.11 &  0.40 &    105 &      8 & BAL \\
SDSSJ104637.30+075318.7 & 1.3635 & 4 & 26/03/14 &   0.04 &   0.09 &   0.10 &  0.07 &  0.08 &     34 &     25 & RadioS \\
SDSSJ104733.16+052454.9 & 1.3341 & 4 & 26/03/14 &   0.04 &  -0.19 &   0.19 &  0.07 &  0.18 &    142 &     11 & \\
SDSSJ105010.05+043249.1 & 1.2158 & 4 & 26/03/14 &  -2.46 &  -1.04 &   2.67 &  0.08 &  2.67 &    102 &      1 & RadioS \\
SDSSJ105018.10+052826.4 & 1.3067 & 4 & 26/03/14 &  -0.11 &   0.24 &   0.26 &  0.10 &  0.24 &     57 &     12 & \\
SDSSJ105422.47+033719.3 & 1.2278 & 4 & 26/03/14 &  -0.18 &  -0.28 &   0.33 &  0.11 &  0.31 &    118 &     10 & RadioS \\
SDSSJ105423.26+051909.8 & 1.2785 & 4 & 26/03/14 &   0.13 &   0.22 &   0.26 &  0.09 &  0.25 &     30 &     11 & BAL \\
SDSSJ105512.23+061243.9 & 1.3018 & 4 & 26/03/14 &  -0.61 &  -0.77 &   0.98 &  0.12 &  0.97 &    116 &      4 & \\
SDSSJ105534.66+033028.8 & 1.2495 & 4 & 26/03/14 &  -0.09 &   0.38 &   0.39 &  0.09 &  0.38 &     52 &      7 & \\
SDSSJ105537.63+040520.0 & 1.2619 & 4 & 26/03/14 &  -0.10 &   0.02 &   0.10 &  0.11 &  0.00 &    $-$ &    $-$ & \\
SDSSJ105719.23+045548.2 & 1.3355 & 4 & 26/03/14 &  -0.36 &   0.44 &   0.57 &  0.13 &  0.56 &     65 &      7 & RadioS \\
SDSSJ105821.28+053448.9 & 1.2540 & 4 & 26/03/14 &   0.09 &   0.06 &   0.11 &  0.10 &  0.06 &     15 &     47 & \\
SDSSJ105833.86+055440.2 & 1.3222 & 4 & 25/03/14 &   0.15 &   0.60 &   0.62 &  0.21 &  0.59 &     38 &     10 & \\
SDSSJ110108.00+043849.6 & 1.2516 & 4 & 25/03/14 &   0.52 &   0.66 &   0.84 &  0.10 &  0.83 &     26 &      3 & \\
SDSSJ110412.00+044058.2 & 1.2554 & 4 & 25/03/14 &   0.15 &  -0.08 &   0.17 &  0.11 &  0.14 &    167 &     22 & \\

\end{longtable}
\tablefoot{Column~1 gives the quasar SDSS name, column~2 the redshift
$z$, column~3 the quasar group (1,2,3 = Huge-LQG, 4 = CCLQG; see also
Fig.~\ref{stru}), column~4 the observation date, columns~5-6 the $q$
and $u$ normalized Stokes parameters corrected for the offset angle,
column~7 the polarization degree $p$, column~8 the error on the
polarization degree $\sigma_p \simeq \sigma_q \simeq \sigma_u$,
column~9 the polarization degree $p_0$ debiased according to the
Wardle and Kronberg (\cite{war74}) method (see also Simmons and
Stewart \cite{sim85}), column~10 the polarization position angle
$\theta$, defined between 0\degr\ and 180\degr\ and measured in the
equatorial coordinate system (North = 0\degr\ and East = 90\degr), and
column~11 the error of the polarization angle estimated using
$\sigma_{\theta} = 28.65\degr \, \sigma_p / p_0$ to avoid
underestimation at low signal-to-noise (Wardle and Kronberg
\cite{war74}). Additional characteristics of the targets were
retrieved from the NASA/IPAC Extragalactic Database (NED) and given in
column~12: radio-source (RadioS), broad absorption line (BAL) quasar,
and low-ionization BAL (LoBAL) quasar.}
\end{longtab}
}

\begin{table*}
\caption{The sample of 19 quasars with $p \geq 0.6\%$}
\label{poltab2}
\begin{tabular}{lccrrrrcc}
\hline\hline
Object & $z$ & LQG & $p$ & $\sigma_{p}$ & $\theta$ \ & $\sigma_{\theta}$ & FWHM & $\sigma_{\rm FWHM}$\\ 
&  &  & (\%) & (\%) & ($\degr$) & ($\degr$) & km s$^{-1}$ & km s$^{-1}$\\
\hline 
SDSSJ105421.90+212131.2 & 1.2573 & 1 & 1.04 & 0.08 &  92.6 & 2.2 & 5094 & 214 \\    
SDSSJ105446.73+195710.5 & 1.2195 & 1 & 1.89 & 0.23 &  75.2 & 3.5 & 3256 & 363 \\    
SDSSJ105611.27+170827.5 & 1.3316 & 1 & 1.29 & 0.08 &  44.8 & 1.8 & 6088 & 158 \\    
SDSSJ110016.88+193624.7 & 1.2399 & 1 & 1.14 & 0.23 & 160.4 & 5.9 & 3909 & 348 \\    
SDSSJ104445.03+151901.6 & 1.2336 & 2 & 1.25 & 0.11 & 167.5 & 2.5 & 3254 & 196 \\  
SDSSJ104616.31+164512.6 & 1.2815 & 2 & 1.25 & 0.11 &  86.9 & 2.5 & 2635 & 222 \\   
SDSSJ104859.74+125322.3 & 1.3597 & 2 & 0.72 & 0.13 &  45.6 & 5.3 & 3746 & 397 \\    
SDSSJ104941.67+151824.6 & 1.3390 & 2 & 1.31 & 0.13 & 146.4 & 2.9 & 4034 & 633 \\    
SDSSJ105245.80+134057.4 & 1.3544 & 2 & 1.32 & 0.11 &  30.2 & 2.4 & 5885 & 174 \\    
SDSSJ105442.71+104320.6 & 1.3348 & 2 & 0.73 & 0.11 & 172.8 & 4.4 & 4108 & 269 \\    
SDSSJ105525.68+113703.0 & 1.2893 & 2 & 2.55 & 0.10 &  49.1 & 1.1 & 4443 & 399 \\    
SDSSJ111009.58+075206.8 & 1.2123 & 3 & 1.81 & 0.17 &  34.2 & 2.7 & 5032 & 626 \\    
SDSSJ111802.11+103302.4 & 1.2151 & 3 & 3.97 & 0.10 & 142.4 & 0.7 & 6900 &1256 \\   
SDSSJ104116.79+035511.4 & 1.2444 & 4 & 1.55 & 0.11 &  99.7 & 2.0 & 2195 & 296 \\   
SDSSJ104225.63+035539.1 & 1.2293 & 4 & 0.69 & 0.08 &  23.2 & 3.3 & 5182 & 380 \\    
SDSSJ105010.05+043249.1 & 1.2158 & 4 & 2.67 & 0.08 & 101.5 & 0.9 & 2703 & 190 \\   
SDSSJ105512.23+061243.9 & 1.3018 & 4 & 0.98 & 0.12 & 115.9 & 3.5 & 3381 & 299 \\   
SDSSJ105833.86+055440.2 & 1.3222 & 4 & 0.62 & 0.21 &  37.8 &10.3 & 5167 & 410 \\    
SDSSJ110108.00+043849.6 & 1.2516 & 4 & 0.84 & 0.10 &  25.7 & 3.4 & 4823 & 269 \\   
\hline
\end{tabular}
\tablefoot{Column~1 gives the quasar SDSS name, column~2 the redshift
$z$, column~3 the quasar group (Fig.~\ref{stru}), columns~4 and 5 the
polarization degree $p$ ans its error $\sigma_p$, columns~6 and 7 the
polarization position angle $\theta$ and its error $\sigma_{\theta}$,
columns~8 and 9 the MgII emission line FWHM and its error from Shen et
al. (\cite{she11}).}
\end{table*}

\begin{figure}[t]
\resizebox{\hsize}{!}{\includegraphics*{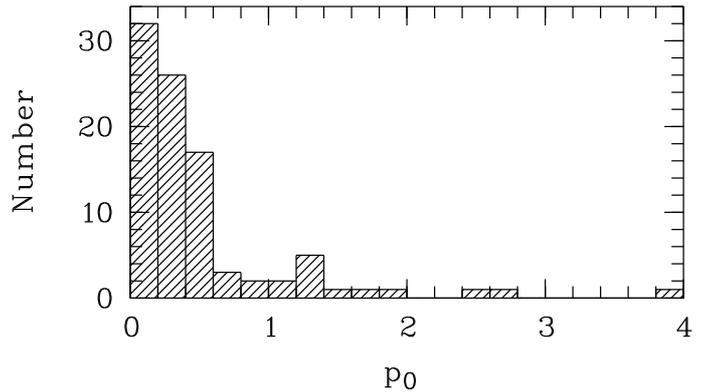}}
\caption{The distribution of the debiased polarization
degree $p_0$ (in \%) measured for the sample of 93 quasars.}
\label{fig1}
\end{figure}

All observations were obtained using the FORS2 V$\_$high filter
($\lambda_0$ = 555 nm, FWHM = 123 nm). Data reduction and measurements
were performed as detailed in Sluse et al. (\cite{slu05}). The
instrumental polarization was checked using the unpolarized stars
WD0752$-$676 and WD1615$-$154 (Fossati et al. \cite{fos07}) and found
to be $p = 0.05 \pm 0.06 \%$, which is consistent with
zero\footnote{We also observed HD~64299 which turned out to be
polarized with $p$ = 0.17 $\pm$ 0.04 \%, in agreement with Masiero et
al. (\cite{mas07}).}.  Note that we did not use field stars to
estimate the instrumental polarization because of spurious off-axis
polarization in FORS1/2 (Patat and Romaniello \cite{pat06}).  To fix
the zero-point of the polarization position angle, polarized standard
stars have been observed: NGC~2024-1, Ve~6-23, CD-28\degr 13479,
HD~316232, BD-14\degr 922 (Fossati et al. \cite{fos07}). The offset
--to subtract from the raw polarization angle-- was determined to be
2.5\degr$\pm$0.5\degr\ in the V$\_$high filter.

\begin{figure}[t]
\resizebox{\hsize}{!}{\includegraphics*{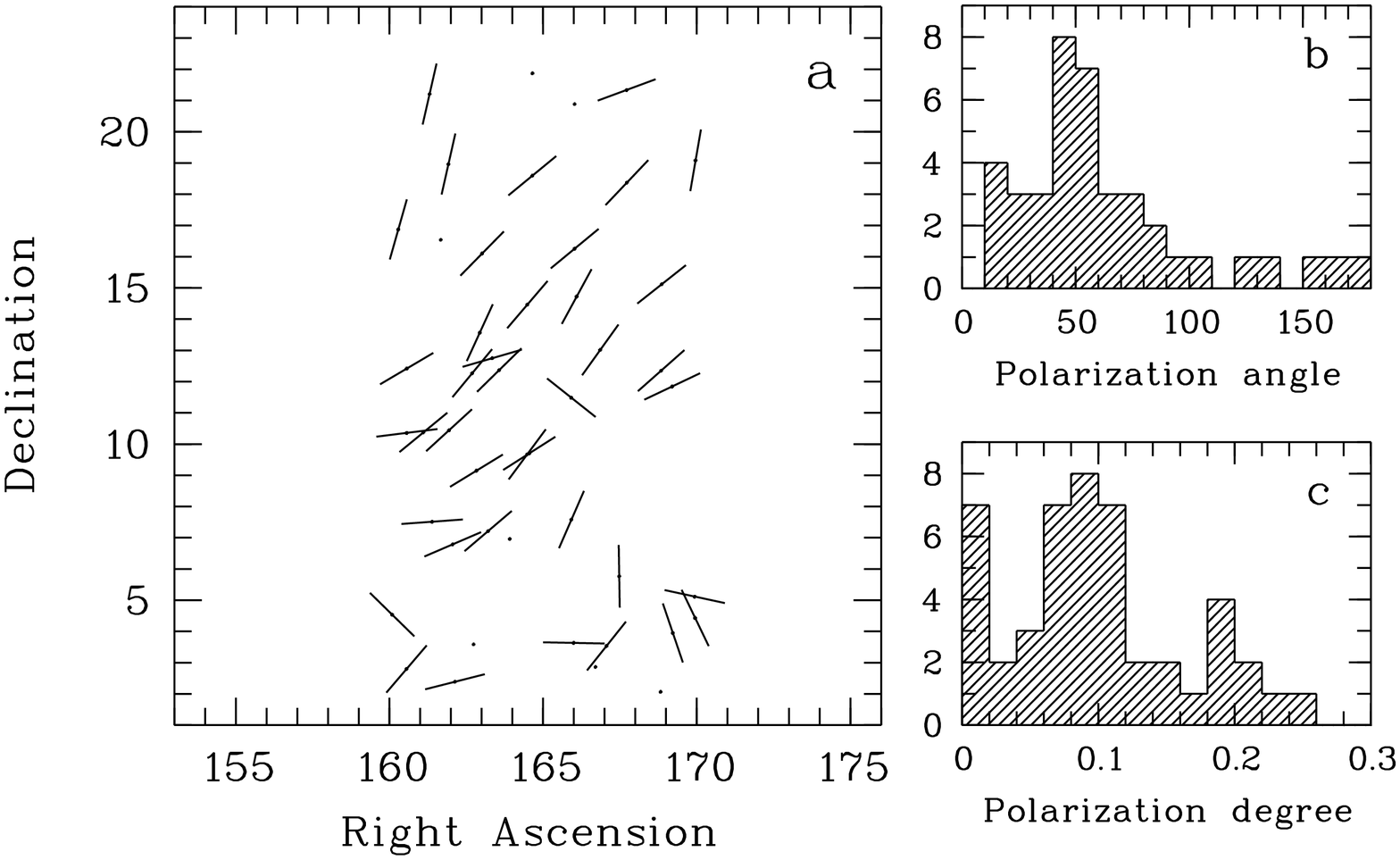}}
\caption{The interstellar polarization in the region of the sky
corresponding to the quasar large-scale structure under study (data
from Berdyugin et al. \cite{ber14}).  (a) Map of polarization vectors;
right ascensions and declinations are in degree; the length of the
polarization vectors is arbitrary.  (b) Distribution of polarization
angles (in degree). (c) Distribution of polarization degrees (in \%).}
\label{maps}
\end{figure}

\begin{figure}[t]
\resizebox{\hsize}{!}{\includegraphics*{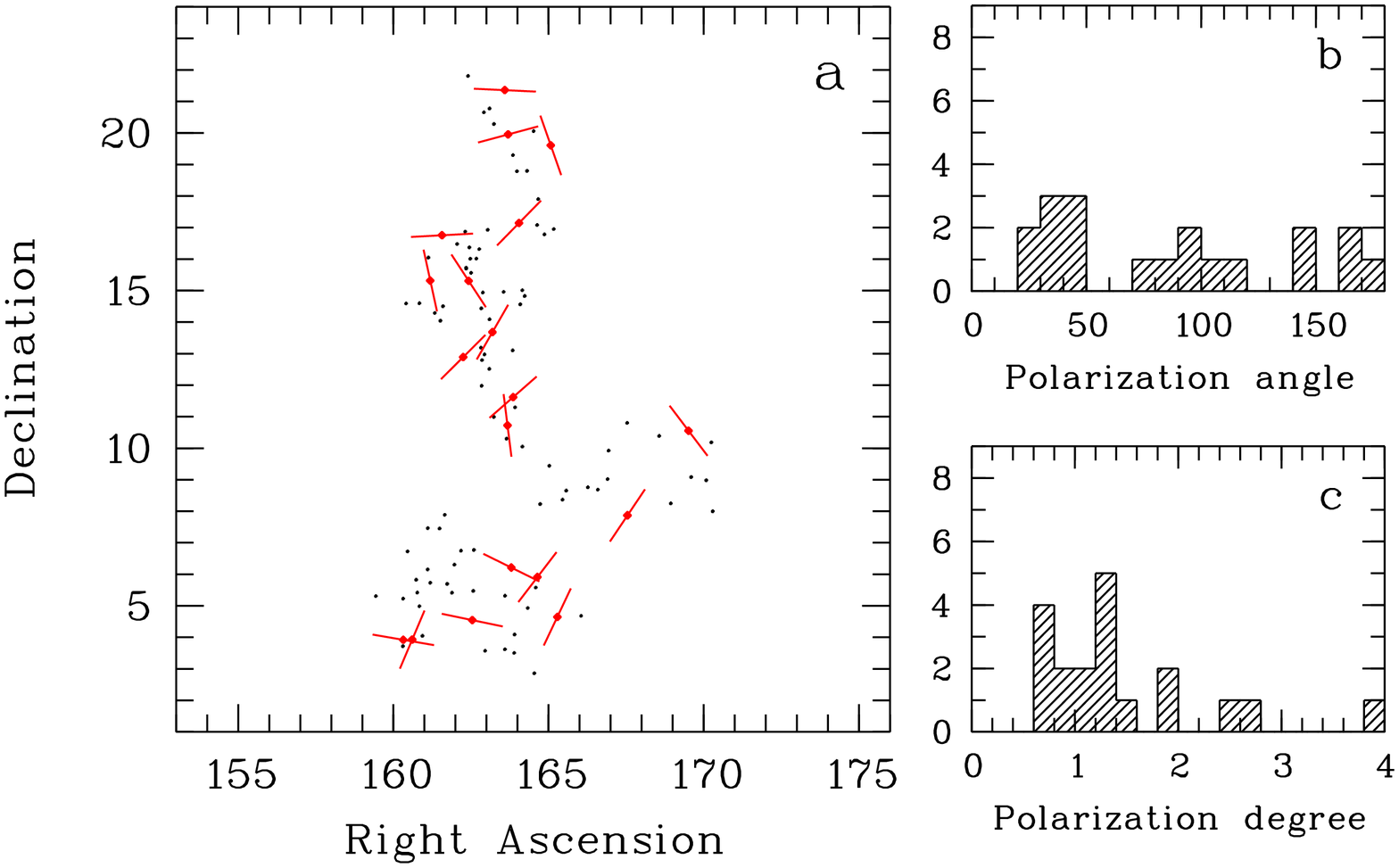}}
\caption{The polarization of the 19 quasars with $p\geq 0.6 \%$. (a)
Map of polarization vectors over the large-scale structure; right
ascensions and declinations are in degree; the length of the
polarization vectors is arbitrary. (b) Distribution of polarization
angles (in degree). (c) Distribution of polarization degrees (in \%).}
\label{map6}
\end{figure}

The linear polarization of all 73 quasars of the Huge-LQG and of 20
out of the 34 quasars of the CCLQG has been obtained, i.e., for a
total of 93 quasars. Table~\ref{poltab} summarizes the measurements.
The error on the polarization degree is between 0.06\% and 0.23\%,
with a mean value of 0.12\%. The distribution of the debiased
polarization degree is illustrated in Fig.~\ref{fig1}. It shows a peak
near the null value, in agreement with other polarization measurements
of radio-quiet non-BAL quasars (Berriman et al. \cite{ber90},
Hutsem\'ekers et al. \cite{hut98b}).  All objects are at galactic
latitudes higher than 50\degr\ which minimizes contamination by
interstellar polarization.  In this region of the sky, the
interstellar polarization is around $p_{\rm is} \simeq 0.1\%$ with a
peak near 50\degr\ (Fig.~\ref{maps}).  As in Papers I-III, we consider
that polarization is essentially intrinsic to the quasar when $p \geq
0.6 \%$ (Berriman et al. \cite{ber90}, Hutsem\'ekers et
al. \cite{hut98b}, Sluse et al. \cite{slu05}). Out of 93 quasars, 19
have $p \geq 0.6 \%$. Their properties are given in
Table~\ref{poltab2}. For these 19 polarized quasars, $\sigma_{\theta}
\leq 10\degr$ with an average value around 3\degr.

\section {Analysis of polarization alignments}
\label{sec:ana}

In Fig.~\ref{map6} we show a map of the quasar polarization vectors
over the LQG structure.  The map does not show any evidence for
coherent orientations or alignments. The distribution of the
polarization angles is flat, compatible with random orientations and
with no contamination by interstellar polarization.

\begin{figure}[t]
\resizebox{\hsize}{!}{\includegraphics*{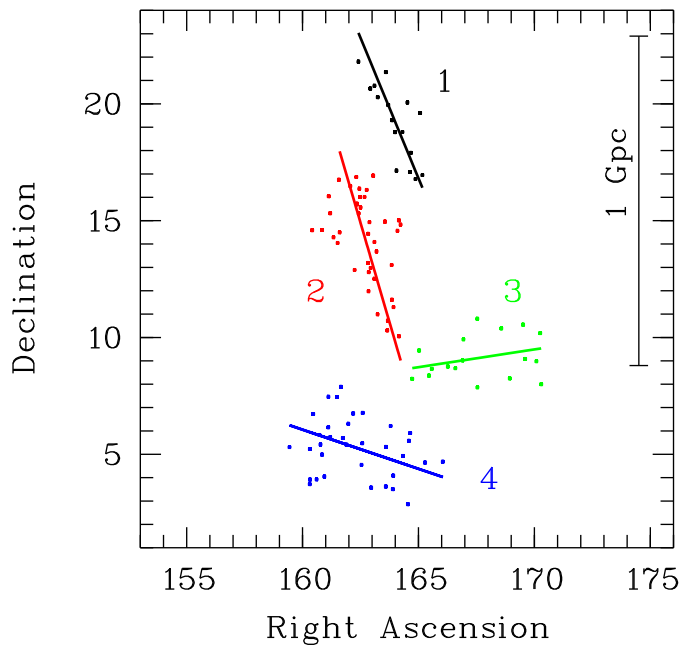}}
\caption{The quasar groups and their orientations on the sky. Right
ascensions and declinations are in degree. The superimposed lines
illustrate the orientations of the four groups labelled 1, 2, 3,
4. The comoving distance scale at redshift $z$ = 1.3 is indicated
assuming a flat Universe with $H_0$ = 70 km s$^{-1}$ Mpc$^{-1}$ and
$\Omega_m$ = 0.27.}
\label{stru}
\end{figure}

In order to compare the quasar polarization angles to the direction of
the local structures, we consider four structures for which we
determine a mean orientation, as illustrated in Fig.~\ref{stru}.
Group 4 is the CCLQG defined in Clowes et al. (\cite{clo12}).  The
Huge-LQG is divided in groups denoted 1, 2 and 3.  Group 3 corresponds
to the ``branch'' set of 17 quasars identified by Clowes et
al. (\cite{clo13}).  The large vertical part of the Huge-LQG is then
separated into groups 1 and 2.  The mean projected direction of the
structures is determined by an orthogonal regression in right
ascension, declination (Isobe et al. \cite{iso90}).  For groups 1, 2,
3, 4, we measure the position angles PA = 157\degr, 164\degr, 81\degr,
and 109\degr , respectively.  We estimate the acute angle between the
quasar polarization vectors and the PA of the structures to which they
belong using $\Delta \theta = \min \left( |{\rm PA} - \theta\,|\, , \,
180\degr - |{\rm PA} - \theta\,| \right) $.

\begin{figure}[t]
\resizebox{\hsize}{!}{\includegraphics*{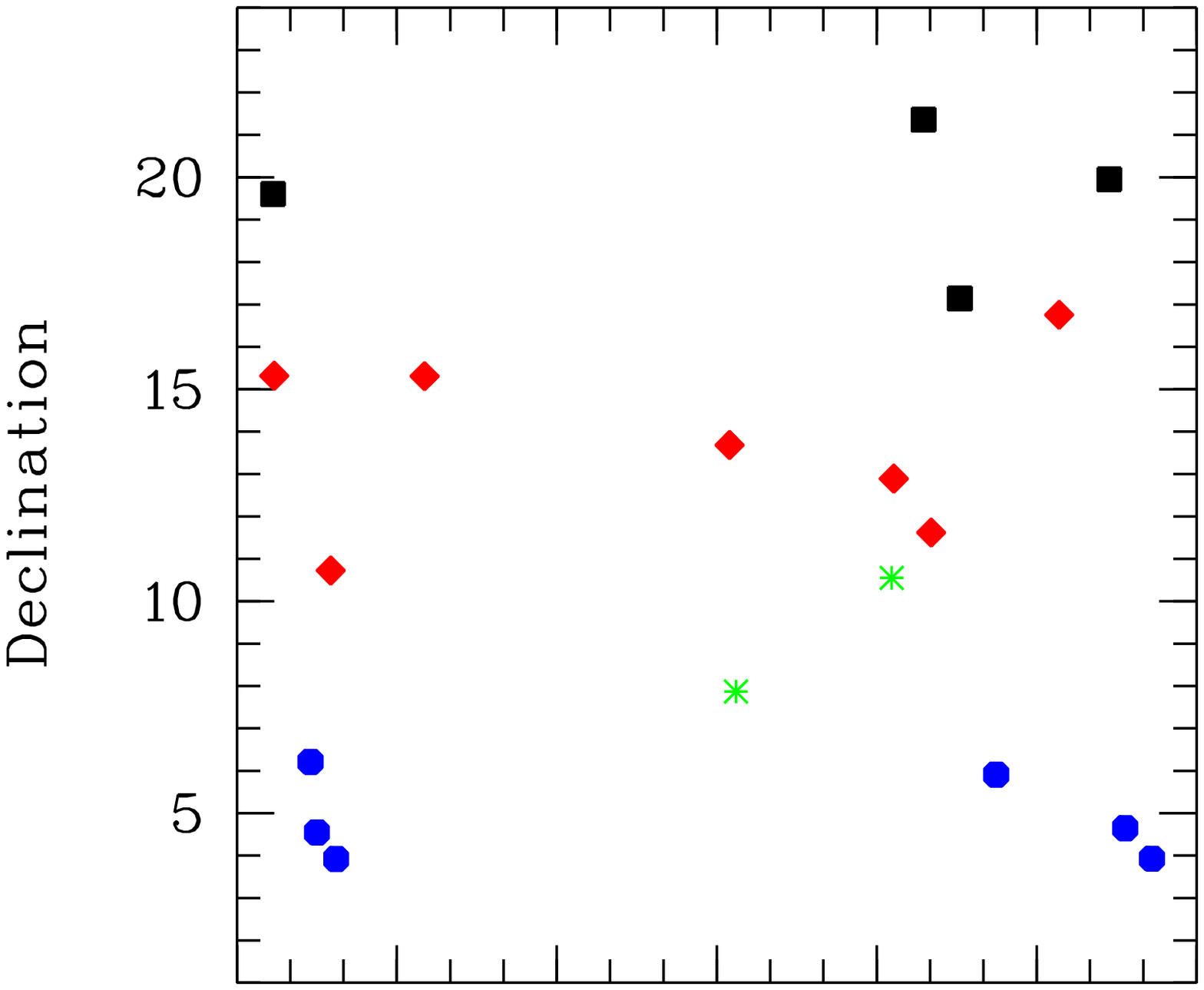}}\\%
\resizebox{\hsize}{!}{\includegraphics*{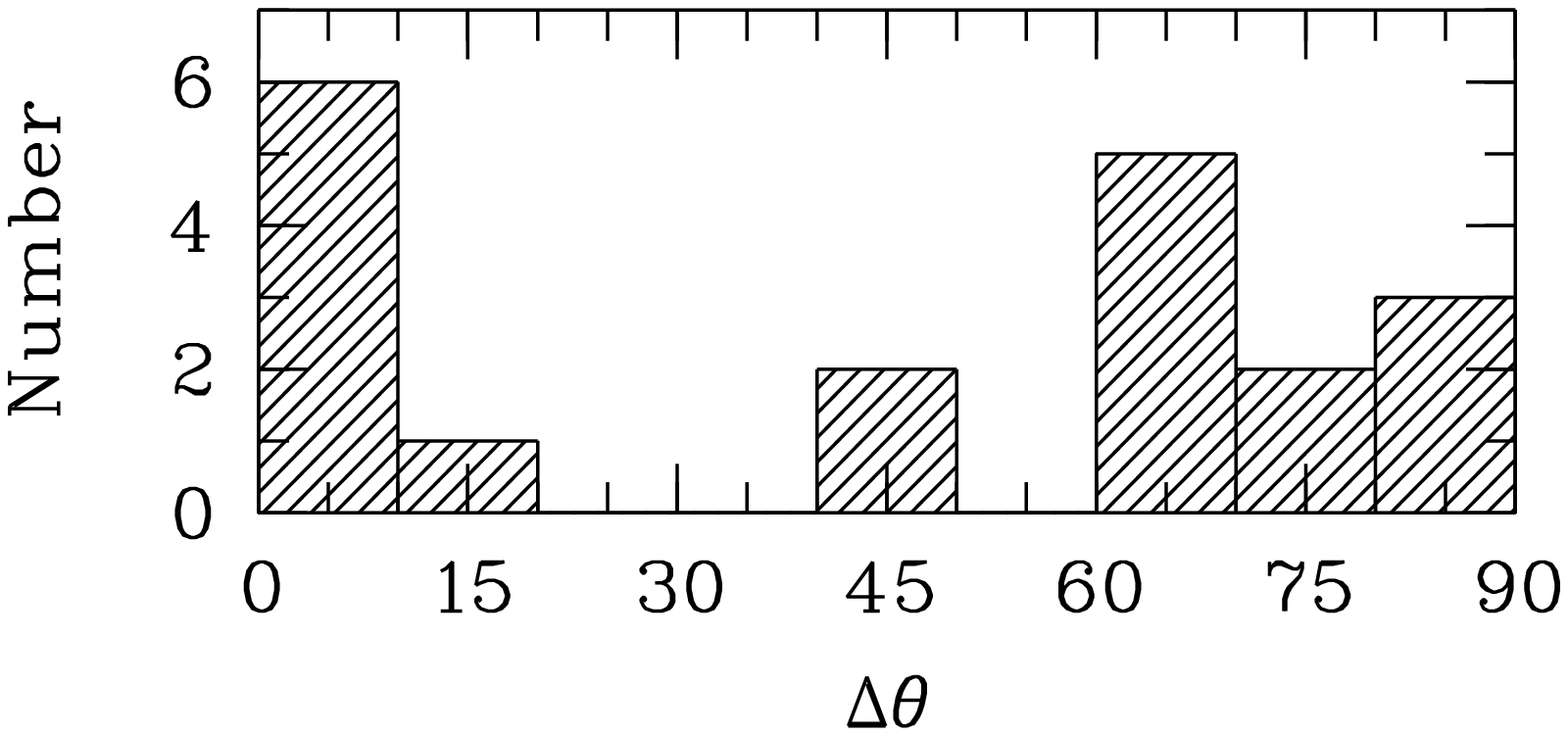}}
\caption{Bottom: The distribution of the acute angle $\Delta \theta$
(in degree) between quasar polarizations and the orientation of their
host large-scale structure. Top: $\Delta \theta$ is plotted against
the object declination (in degree) to illustrate the behavior of the
different quasar groups (1: squares, 2: lozenges, 3: asterisks, 4:
hexagons; colors as in Fig.~\ref{stru}).}
\label{dteta6}
\end{figure}

The distribution of $\Delta \theta$ is illustrated in
Fig.~\ref{dteta6}. It shows a bimodal distribution, with both
alignments ($\Delta \theta \simeq 0\degr$) and anti-alignments
($\Delta \theta \simeq 90\degr$) in each quasar group (except group
3). The cumulative binomial probability to have 9 or more quasars in
the first and the last bins is $P_{\rm bin}$ = 1.4\%.  The Kuiper test
(Arsham \cite{ars88}, Fisher \cite{fis93}) gives a probability $P_K$ =
1.6\% that the observed distribution is drawn from an uniform
distribution. These results are robust if we consider the 28
quasars with $p \geq$ 0.5\% (in this case $P_{\rm bin}$ = 1.2\% and
$P_K$ = 1.0\%).

\begin{figure}[t]%
\resizebox{\hsize}{!}{\includegraphics*{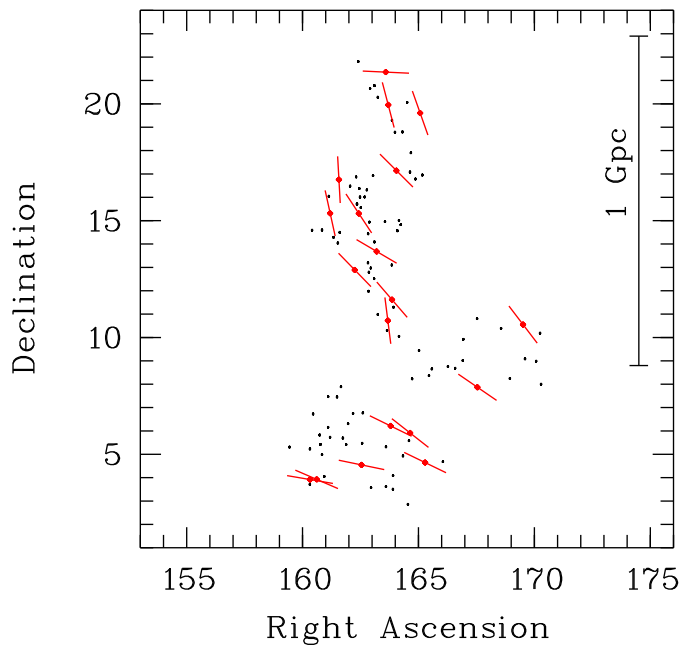}}%
\caption{The polarization vectors of the 19 quasars with $p\geq 0.6\%$
are superimposed on the large-scale structure after rotation of the
polarization angles according to $\tilde{\theta} =
\rm{mod}(\theta,90\degr) +90\degr$.  A clear correlation is seen but
we nevertheless caution against exaggerated visual impression since
polarization angles are now in the range [90\degr -- 180 \degr]. Right
ascensions and declinations are in degree. The comoving distance scale
is indicated as in Fig.~\ref{stru}.}
\label{maprot6}
\end{figure}

A bimodal distribution of $\Delta \theta$ is exactly what we expect if
the quasar morphological axes are related to the orientation of the
host large-scale structures. Indeed, the polarization of type 1 AGN is
usually either parallel or perpendicular to the AGN accretion disk
axis depending on the inclination with respect to the line of sight
(e.g., Smith et al. \cite{smi04}).  We may assume that higher
luminosity AGN (quasars) behave similarly. In Fig.~\ref{maprot6}, the
quasar polarization angles modified according to $\tilde{\theta} =
\rm{mod}(\theta,90\degr) +90\degr$ are plotted over the LQG structure,
unveiling a remarkable correlation. We stress that such a behavior
cannot be due to contamination by interstellar polarization which
would align all polarizations similarly.

To quantify the significance of this correlation independently of the
shape of the host structure, we use the Andrews and Wasserman $Z_c$
statistical test (Bietenholz \cite{bie86}, Paper~I).  This test is
best suited to small samples since it does not involve angle
dispersion. The idea of the Andrews \& Wasserman test is to compute
for each object $i$, the mean direction $\bar{\theta}_{i}$ of its
$n_{v}$ nearest neighbours, and to compare this local average to the
polarization angle of the object $i$, $\theta_{i}$. If angles are
correlated to positions, one expects, on the average, $\theta_{i}$ to
be closer to $\bar{\theta}_{j=i}$ than to $\bar{\theta}_{j \neq i}$.
As a measure of the closeness of $\theta_{i}$ and $\bar \theta_{j}$,
one uses D$_{i,j}$ = ${\bf y}_{i} {\bf .} {\bf \bar Y}_{j}$, where
${\bf y}_{i}$ is the normalized polarization vector of object $i$ and
${\bf \bar Y}_{j}$ the normalized resultant polarization vector of the
$n_{v}$ neighbours of object $j$, excluding $j$.  Then $Z_i$ is
computed by ranking D$_{i,j=i}$ among the D$_{i,j=1,n}$ and the final
statistics $Z_c$ is obtained by averaging the $Z_i$ over the sample of
$n$ objects.  $Z_{c}$ is expected to be significantly larger than zero
when the polarization angles are not randomly distributed over
positions.  To make the test independent of the coordinate system,
polarization vectors can be parallel transported before computing the
resultant polarization vectors (Jain et al. \cite{jai04}).  Here, the
polarization vectors are computed using ${\bf y} = (\cos\Theta,
\sin\Theta)$ with $\Theta = 4 \, \rm{mod}(\theta,90\degr)$ instead of
$\Theta = 2 \theta = 2 \, \rm{mod}(\theta,180\degr)$ to test for
either alignments or anti-alignments (i.e., dealing with 4-axial data
instead of 2-axial data, Fisher \cite{fis93}).  To estimate the
statistical significance, 10$^5$ samples of 19 angles were created
through Monte-Carlo simulations either by shuffling the measured
angles over positions, or by randomly generating them in the
[0\degr,180\degr] range (Press et al. \cite{pre92}).  The significance
level (S.L.) of the test is finally computed as the percentage of
simulated configurations for which $Z_c \geq Z_c^{\star}$ where
$Z_c^{\star}$ is the measured sample statistics.  Since all quasars
are in a limited redshift range, we only consider their angular
positions on the sphere.

\begin{figure}[t]
\resizebox{\hsize}{!}{\includegraphics*{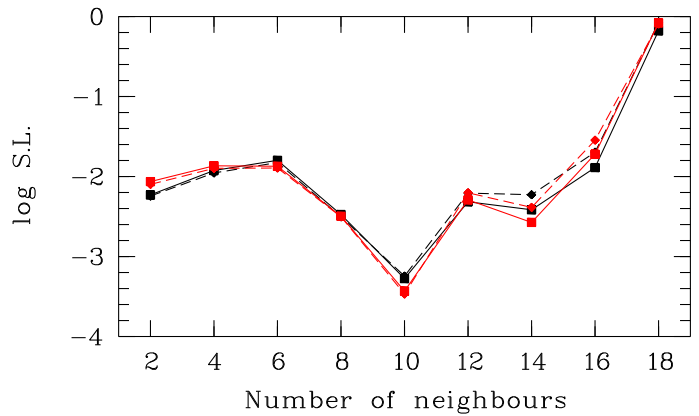}}
\caption{The logarithm of the significance level (S.L.) of the $Z_c$
test applied to the sample of 19 polarized quasars, as a function of
the number of nearest neighbours. The solid line refers to simulations
obtained by shuffling angles over positions while the dashed line
refers to simulations obtained by randomly generating angles. The
statistics are computed with (in red) and without (in black) parallel
transport of the polarization vectors.}
\label{sl6}
\end{figure}

The significance level of the $Z_c$ test is illustrated in
Fig.~\ref{sl6}. It shows that the probability that the polarization
angles are randomly distributed over positions is smaller than
1\%. The effect is stronger (S.L. $<$ 0.1\%) when the mean orientation
is computed with 10 nearest neighbours, i.e., roughly half of the
sample. This number corresponds to a mean comoving distance of $\sim$
550 Mpc, in agreement with the trend seen in Fig.~\ref{maprot6}. 
Parallel transport has little effect since all quasars lie close to
each other and to the equator.  We emphasize that a deviation from
uniformity is only detected when using $4\, \rm{mod}(\theta,90\degr)$
in the $Z_c$ test and not when using $2\, \rm{mod}(\theta,180\degr)$,
which means that purely parallel or perpendicular alignments are not
seen (Fig.~\ref{map6}). If we consider the 28 quasars with $p \geq$
0.5\%, a similar curve is obtained with the minimum shifted to $n_v$ =
20, which corresponds to a mean comoving distance of $\sim$ 650 Mpc.

\begin{figure}[t]
\resizebox{\hsize}{!}{\includegraphics*{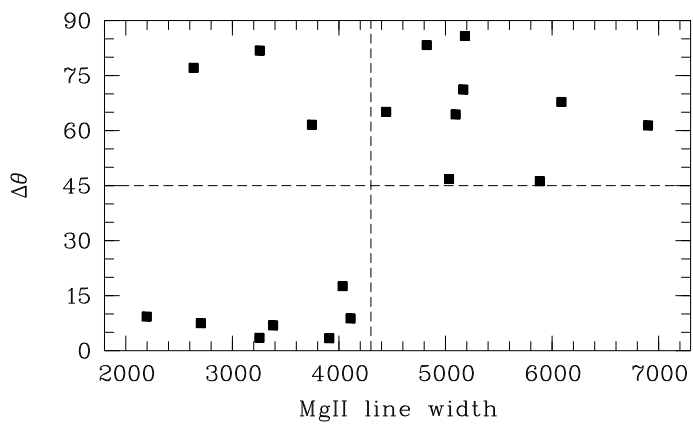}}
\caption{The angle $\Delta \theta$ (in degree) between quasar
polarizations and the orientation of their host large-scale structures
as a function of the MgII emission line width (FWHM in km~s$^{-1}$).}
\label{tewi6}
\end{figure}

\begin{figure}[t]
\resizebox{\hsize}{!}{\includegraphics*{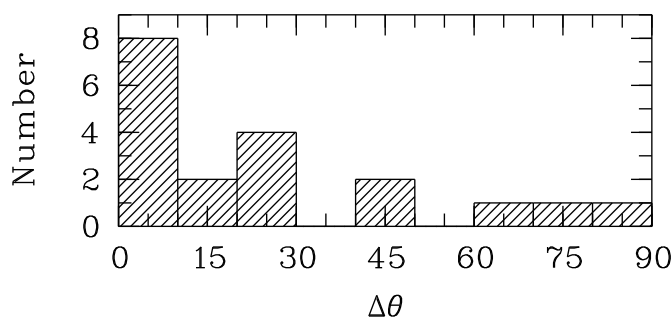}}%
\caption{The distribution of the acute angle $\Delta \theta$ (in
degree) between quasar polarizations and the orientation of their host
large-scale structure after rotating by 90\degr\ the polarization
angles of objects with MgII emission line widths larger than 4300 km
s$^{-1}$.}
\label{histrot6}
\end{figure}

Since the width of low-ionization emission lines (H$\beta$, MgII)
observed in quasar spectra correlates with the object inclination with
respect to the line of sight (Wills and Brown \cite{wil86}, Brotherton
\cite{bro96}, Jarvis and McLure \cite{jar06}, Decarli et
al. \cite{dec08}), we plot in Fig.~\ref{tewi6} the angle $\Delta
\theta$ as a function of the quasar MgII emission line width (FWHM
from Shen et al. \cite{she11}).  We see that most objects with
polarization perpendicular to the host structure ($\Delta \theta >
45\degr$) have large emission line widths while all objects with
polarization parallel to the host structure ($\Delta \theta <
45\degr$) have small emission line widths.  A two-sample
Kolmogorov-Smirnov test indicates that there is a probability of only
1.4\% that quasars with either perpendicular or parallel polarizations
have emission line widths drawn from the same parent population.
Quasars seen at higher inclinations\footnote{Face-on: $i =
0\degr$. Edge-on: $i = 90\degr$} generally show broader low-ionization
emission lines, in agreement with line formation in a rotating disk
(Wills and Brown \cite{wil86}, Jarvis and McLure \cite{jar06}, Decarli
et al. \cite{dec08}). The relation seen in Fig.~\ref{tewi6} thus
supports our hypothesis that the polarization of quasars is either
parallel or perpendicular to the host structure depending on their
inclination. When rotating by 90\degr\ the polarization angles of
objects with MgII emission line widths larger than 4300 km s$^{-1}$, a
stronger alignment is seen (Fig.~\ref{histrot6}). The Kuiper test
gives a probability $P_K$ = 0.5\% that the observed distribution is
drawn from an uniform distribution. But this value should be seen with
caution since the cut at 4300 km s$^{-1}$ is arbitrary. On the other
hand, it should be emphasized that the emission line width does not
only depend on inclination but also on the mass of the central black
hole if the rotating disk is virialized. Quasars with lower black hole
mass will have narrower emission lines whatever their inclination so
that some of them may still appear anti-aligned in Fig.~\ref{histrot6}.

Since objects seen at higher inclinations preferentially show
polarization perpendicular to their axes (Smith et al. \cite{smi04}),
we finally infer that quasar spin axes should be predominantly
parallel to the orientation of the structures to which they belong.

\section {Conclusions}
\label{sec:conclu}

We have measured the polarization of 93 quasars belonging to
large-scale quasar groups. 19 quasars out of 93 are significantly
polarized with $p\geq 0.6 \%$.

We found that quasar polarization vectors are either parallel or
perpendicular to the large-scale structures to which they belong, and
correlated to the polarization vectors of their neighbours. The
probability that these results can be attributed to a random
distribution of polarization angles is of the order of 1\%.  Such a
behavior cannot be due to contamination by interstellar
polarization. Our results are robust if we consider $p \geq 0.5 \% $
instead of $p \geq 0.6 \%$, or if we subtract a systematic $p_{\rm is}
= 0.1\% $ at $\theta_{\rm is} = 50\degr $ to simulate the correction
of a possible contamination by interstellar polarization
(Fig.~\ref{maps}).

Assuming that quasar polarization is either parallel or perpendicular
to the accretion disk axis as a function of inclination, as
observed in lower luminosity AGN, and considering that broader
emission lines originate from quasars seen at higher inclinations, we
inferred that quasar spin axes are likely parallel to their host
large-scale structures.

Galaxy spin axes are known to align with large-scale structures such
as cosmic filaments (e.g., Tempel and Libeskind \cite{tem13}, Zhang et
al. \cite{zha13}, and references therein).  Till now, such alignments
are detected up to redshift $z \sim 0.6$ at scales $\lesssim 100$ Mpc
(Li et al. \cite{lic13}).  Detailed interpretations remain complex
because the link between galaxy and halo spin axes is not
straightforward, and because the strength and orientation of the
alignments depend on several factors, in particular the mass of the
halo and the cosmic history (e.g., Hahn et al. \cite{hah10}, Trowland
et al. \cite{tro13}, Dubois et al. \cite{dub14}).  We have found that
quasar accretion disk axes are likely parallel to the large-scale
structures to which they belong over Gpc scales at redshift $z \sim$
1.3 , i.e., one order of magnitude bigger than currently known galaxy
alignments.  Although the scales involved are much larger, we may
assume that similar mechanisms can explain alignments of quasar and
galaxy axes with their host large-scale structure, keeping in mind
that polarization-related quasar regions (accretion disk, jet,
scattering region) are not necessarily well aligned with the stellar
component of the host galaxy (Borguet et al. \cite{bor08}, Hopkins et
al. \cite{hop12}), and that quasars, more prone to strong feedback
mechanisms, can have a different cosmic history (Dubois et
al. \cite{dub14}).

Since coherent orientations of quasar polarization vectors, and then
quasar axes, are found on scales larger than 500 Mpc, our results
might also provide an explanation to the very large-scale polarization
alignments reported in Papers~I-III. In this case those alignments
would be intrinsic and not due to a modification of the polarization
along the line of sight.  The existence of correlations in quasar axes
over such extreme scales would constitute a serious anomaly for the
cosmological principle.

\begin{acknowledgements}

We thank the referee, E. Tempel, for constructive remarks which helped
to improve the paper.  We thank B. Borguet, R. Cabanac, J.R. Cudell,
H. Lamy and A. Payez for discussions and V. Ivanov for his efficient
help during the observations.  D.H. and L.B. are respectively Senior
Research Associate and Research Assistant at the F.R.S.-FNRS.
D.S. acknowledges support from the Deutsche Forschungsgemeinschaft,
reference SL172/1-1.  This work was supported by the Fonds de la
Recherche Scientifique - FNRS under grant 4.4501.05.  This research
has made use of the NASA/IPAC Extragalactic Database (NED) which is
operated by the Jet Propulsion Laboratory, California Institute of
Technology, under contract with the National Aeronautics and Space
Administration.

\end{acknowledgements}


\begin{thebibliography}{999}
\bibitem[2012]{aga12} Agarwal, N., Aluri, P.K., Jain, P., Khanna, U., Tiwari, P. 2012, EPJC, 72, 1928
\bibitem[1988]{ars88} Arsham H. 1988, J. Appl. Stat. 15, 131
\bibitem[2014]{ber14} Berdyugin, A., Piirola, V., Teerikorpi, P. 2014, \aap, 561, A24
\bibitem[1990]{ber90} Berriman, G., Schmidt, G.D., West, S.C., Stockman, H.S. 1990, \apjs, 74, 869
\bibitem[1986]{bie86} Bietenholz, M.F. 1986, \aj, 91, 1249
\bibitem[2008]{bor08} Borguet, B., Hutsem\'ekers, D., Letawe, G., Letawe, Y., Magain, P. 2008, \aap, 478, 321
\bibitem[1996]{bro96} Brotherton, M.S. 1996, \apjs, 102, 1 
\bibitem[2012]{clo12} Clowes, R.G., Campusano, L.E., Graham, M.J., S\"ochting, I.K. 2012, \mnras, 419, 556
\bibitem[2013]{clo13} Clowes, R.G., Harris, K.A., Raghunathan, S., et al. 2013, \mnras, 429, 2910
\bibitem[2005]{das05} Das, S., Jain, P., Ralston, J.P., Saha, R. 2005, \jcap, 06, 002
\bibitem[2008]{dec08} Decarli, R., Labita, M., Treves, A., Falomo, R. 2008, \mnras, 387, 1237 
\bibitem[2014]{dub14} Dubois, Y., Pichon, C., Welker, C., et al. 2014, \mnras, 444, 1453
\bibitem[2014]{ein14} Einasto, M., Tago, E., Lietzen, H., et al., 2014, \aap, 568, A46 
\bibitem[1993]{fis93} Fisher, N.I. 1993, Statistical Analysis of Circular Data, Cambridge University Press, Cambridge
\bibitem[2007]{fos07} Fossati, L., Bagnulo, S., Mason, E., Landi Degl’Innocenti, E. 2007, ASPC, 364, 503 
\bibitem[2010]{hah10} Hahn, O., Teyssier, R., Marcella Carollo, C. 2010, \mnras, 405, 274
\bibitem[2012]{hop12} Hopkins, P.F., Hernquist, L., Hayward, C.C. 2012, \mnras, 425, 1121
\bibitem[1998]{hut98} Hutsem\'ekers, D. 1998, \aap, 332, 410 (Paper I)
\bibitem[1998b]{hut98b} Hutsem\'ekers, D., Lamy, H., Remy, M. 1998, \aap, 340, 371
\bibitem[2001]{hut01} Hutsem\'ekers, D., Lamy, H. 2001, \aap, 367, 381  (Paper II)
\bibitem[2005]{hut05} Hutsem\'ekers, D., Lamy, H., Cabanac, R., Sluse, D. 2005, \aap, 441, 915 (Paper III)
\bibitem[2010]{hut10} Hutsem\'ekers, D., Borguet, B., Sluse, D., Cabanac, R., Lamy, H. 2010,  \aap, 520, L7
\bibitem[1990]{iso90} Isobe, T., Feigelson, E., Akritas, M.G., Babu, G.J. 1990, \apj, 364, 104 
\bibitem[2004]{jai04} Jain, P.,  Narain, G., Sarala, S. 2004, \mnras, 347, 394
\bibitem[2006]{jar06} Jarvis, M.J., McLure, R.J. 2006, \mnras, 369, 182 
\bibitem[2013]{lic13} Li, C., Jing, Y.P., Faltenbacher, A., Wang, J. 2013, \apj, 770, L12
\bibitem[2007]{mas07} Masiero, J., Hodapp, K., Harrington, D., Lin, H. 2007,  \pasp, 119, 1126
\bibitem[2013]{nad13} Nadathur, S. 2013, \mnras, 434, 398
\bibitem[2006]{pat06} Patat, F., Romaniello, M. 2006, \pasp, 118, 146
\bibitem[2011]{pay11} Payez, A., Cudell, J. R., Hutsemékers, D. 2011, \prd, 84, 085029
\bibitem[2014]{pel14} Pelgrims, V., Cudell, J.R. 2014, \mnras, 442, 1239
\bibitem[1992]{pre92} Press, W.H., Teukolsky, S.A., Vetterling, W.T., Flannery, B.P., 1992, Numerical Recipes. The Art of Scientific Computing, Cambridge University Press, Cambridge
\bibitem[2011]{she11} Shen, Y., Richards, G.T., Strauss, M.A. 2011, \apjs, 194, 45
\bibitem[2013]{shu13} Shurtleff, R., 2013, arXiv:1311.6118
\bibitem[1985]{sim85} Simmons, J. F. L.,  Stewart, B. G. 1985, \aap, 142, 100
\bibitem[2004]{smi04} Smith, J.E., Robinson, A., Alexander, D.M., et al. 2004, \mnras, 350, 140
\bibitem[2005]{slu05} Sluse, D., Hutsem\'ekers, D., Lamy, H., Cabanac, R., Quintana, H. 2005, \aap, 433, 757
\bibitem[2013]{tem13} Tempel, E., Libeskind, N.I. 2013, \apj, 775, L42
\bibitem[2013]{tro13} Trowland, H.E., Lewis, G., Bland-Hawthorn, J. 2013, \apj, 762, 72
\bibitem[1974]{war74} Wardle, J. F. C., Kronberg, P. P. 1974, \apj, 194, 249
\bibitem[1986]{wil86} Wills, B.J., Browne, I.W.A. 1986, \apj, 302, 56 
\bibitem[2010]{yad10} Yadav, J.K., Bagla, J.S., Khandai, N. 2010, \mnras, 405, 2009
\bibitem[2013]{zha13} Zhang, Y., Yang, X., Wang, H., et al. 2013, \apj, 779, 160
\end{thebibliography}
\end{document}